\documentclass[12pt]{article}
\usepackage{amsfonts}
\usepackage[utf8]{inputenc}
\usepackage{bm}
\usepackage{amsmath}
\usepackage{epsfig}
\usepackage{enumitem}

\setlist[enumerate]{%
wide =0.5\parindent,
listparindent=0pt%
}

\newcommand{\bmat}{\left(\begin{array}}
\newcommand{\emat}{\end{array}\right)}

\def\gtrsim{\mathrel{\raise.3ex\hbox{$>$\kern-.75em\lower1ex\hbox{$\sim$}}}}

\def\un{\underline}

\def\s2{\frac{1}{\sqrt2}}

\def\z{\zeta}

\def\ov{\overline}
\def\un{\underline}

\def\Dsl{\,\raise.15ex\hbox{/}\mkern-13.5mu D}

\def\be{\begin{equation}}
\def\ee{\end{equation}}
\def\bea{\begin{eqnarray}}
\def\eea{\end{eqnarray}}

\newcommand{\nn}{\nonumber}

\begin{document}


\pagestyle{plain}

\makeatletter
\@addtoreset{equation}{section}
\makeatother
\renewcommand{\theequation}{\thesection.\arabic{equation}}
\pagestyle{empty}
\begin{center}
\ \

\vskip .5cm

\LARGE{\LARGE\bf Gauged Double Field Theory as an $L_{\infty}$ algebra \\[10mm]}
\vskip 0.3cm

\large{Eric Lescano$^{a,c}$ and Mart\'in Mayo$^{b,c,d}$
 \\[6mm]}

{\small  $^a$ Instituto de Astronom\'ia y F\'isica del Espacio (IAFE-UBA)\\ [.01 cm]}
{\small\it Ciudad Universitaria, Pabell\'on IAFE, 1428 Buenos Aires, Argentina\\ [.1 cm]}
{\small  $^b$ G. F\'isica CAB-CNEA,
\\ [.01 cm]}
{\small\it Centro At\'omico Bariloche, Av. Bustillo 9500, Bariloche, Argentina\\ [.1 cm]}
{\small  $^c$ Consejo Nacional de Investigaciones Cient\'ificas y T\'ecnicas (CONICET)\\ [.1 cm]}
{\small  $^d$ Max-Planck-Institut fur Physik (Werner-Heisenberg-Institut), \\ [.1 cm]}
{\small\it Fohringer Ring 6, 80805 Munchen, Germany\\ [.3 cm]}

{\small \verb"elescano@iafe.uba.ar, martin.mayo@ib.edu.ar"}\\[1cm]

\small{\bf Abstract} \\[0.5cm]\end{center}
 
$L_{\infty}$ algebras describe the
underlying algebraic structure of many consistent classical
field theories. In this work we analyze the algebraic structure of Gauged Double Field Theory in the generalized flux formalism. The symmetry transformations consist of a generalized deformed Lie derivative and double Lorentz transformations. We obtain all the non-trivial products in a closed form considering a generalized Kerr-Schild ansatz for the generalized frame and we include a linear perturbation for the generalized dilaton. The off-shell structure can be cast in an $L_{3}$ algebra and when one considers dynamics the former is exactly promoted to an $L_{4}$ algebra. The present computations show the fully algebraic structure of the fundamental charged heterotic string and the $L^{\rm{gauge}}_{3}$ structure of (Bosonic) Enhanced Double Field Theory.

\newpage


\setcounter{page}{1}
\pagestyle{plain}
\renewcommand{\thefootnote}{\arabic{footnote}}
\setcounter{footnote}{0}

\tableofcontents
\newpage

\section{Introduction}

Lie algebras are vector spaces
equipped with an antisymmetric bracket satisfying the Jacobi identity. Many interesting physical theories can be cast in this language, but others require a suitable generalization of this program. This is the case of Double Field Theory (DFT) \footnote{For reviews check \cite{ReviewDFT} and references therein.} \cite{Siegel} \cite{DFT}, a proposal to incorporate T-duality as a symmetry of a field theory, since it contains a non-trivial Jacobiator and therefore satisfies the Jacobi identity up to homotopy. For this reason its algebraic structure requires a set of brackets defined on a graded vector space satisfying a generalized notion of the Jacobi identities. Such structures are known as $L_{\infty}$ algebras and were initially described in the context of closed string field theory \cite{Zwiebach:1992ie} and, in the mathematics literature, in topology \cite{Topology}.

One way of organizing the algebraic structure of DFT in an $L_\infty$ structure turns from noticing that the Courant algebroids can be cast in this language \cite{Roytenberg:1998vn}, as well as their duality covariant counterparts \cite{Hull:2009zb} \cite{Deser:2016qkw}. Moreover, when dynamics is taken into account the full DFT, written in the generalized metric approach, also fit in an $L_{\infty}$ structure, as described in \cite{Hohm:2017pnh} \cite{Otros}. 

In this work we are interested in Gauged Double Field Theory (GDFT) \cite{GDFT} with $O(D,D+n)$ as global duality group, where $n$ is the dimension of a gauge group. This formalism is a generalization of DFT and requires a frame \cite{frame} or flux formalism \cite{flux} in order to introduce the generalized version of the structure constants $f_{M N P}$. Additionally, the generalized Lie derivative acting on a generic vector $V_{M}$ with $M=0,\dots,2D-1+n$ is consistently deformed, 
\bea
\widehat{\cal L}_\xi V_M = {\cal L}_{\xi} V_{M} + f_{M N P} \xi^{N} V^{P} \, ,
\eea 
and the closure is given by a deformed bracket  
\bea
 [\xi_1, \xi_{2} ]^{ M}_{(C_{f})}=2\xi^{ P}_{[1}\partial_{ P}\xi_{2]}^{ M}-\xi_{[1}^{ N}\partial^{ M}\xi_{2] N}+f_{{ PQ}}{}^{ M} \xi_{1}^{ P} \xi_2^{ Q}\, ,
\eea
which reduces to the C-bracket when the structure constants vanish. As expected, the Jacobiator is also deformed
\be
J(\xi_{1},\xi_{2},\xi_3)^{M} = \frac{3}{2}\partial^{M}(\xi_{[1}^{N} \xi_{2}^{P} \partial_{N} \xi_{3]P} + \frac13 f_{N P Q} \xi_{1}^{N} \xi_{2}^{P} \xi_{3}^{Q}) \, .
\ee
The inclusion of the generalized frame/fluxes introduces a double Lorentz symmetry given by $O(D-1,1)_{L} \times O(1,D-1+n)_{R}$. From a $L_{\infty}$ point of view all these new ingredients enrich the algebraic structure of DFT or, in other words, define the algebraic structure of GDFT. 

The products related to the dynamics of the theory can be cast in a closed form if we restrict our study to a family of theories given by the generalized Kerr-Schild ansatz. This ansatz was introduced in the context of DFT in \cite{KL}, extended to heterotic DFT in \cite{KL2} and \cite{alpha},  and further explore in the context of duality covariant theories in \cite{GKSA}. In this ansatz the perturbation of the generalized exact frame  is given by
\bea
{\cal E}_{M}{}^{\ov A} = E_{M}{}^{\ov A} + \frac12 \kappa E_{M}{}^{\underline B} K_{\underline B} {\bar K}^{\ov A} \, , \nn \\ {\cal E}_{M}{}^{\underline A} = E_{M}{}^{\underline A} - \frac12 \kappa E_{M}{}^{\ov B} {\bar K}_{\ov B} K^{\underline A}\, ,
\label{introan}
\eea
where $K_{\underline A}$ and $\bar K_{\overline A}$ are a pair of generalized null vectors,
\bea
K_{\un A} K^{\un A} & = & \bar K_{\ov A} \bar K^{\ov A} = 0 \, ,    
\eea
and $\kappa$ in (\ref{introan}) is an order parameter. We use $\un A$, $\ov A$ as the flat left and right projections of the $M,N$ indices. The vectors $K_{\un A}$ and $\bar K_{\ov A}$ satisfy the equivalent of a geodesic condition in the context of DFT,
\bea
 K^{\underline A} D_{\underline A} {\bar K}^{\overline B} =
{\bar K}^{\overline A} D_{\overline A} {K}^{\underline B} = 0 \, ,
\eea
where $D_{A}$ is a generalized covariant derivative. The ansatz (\ref{introan}) plus a linear expansion for the generalized dilaton, 
\bea
d = d_{o} + \kappa f
\eea
with $K^{\underline A} E_{\underline A} f = K^{\ov A} E_{\ov A} f = 0$ provide a family of exact solutions in a perturbative framework. In this sense, all the non-trivial products of the $L_{\infty}$ structure of GDFT can be explicitly/exactly computed, as we show. Considering a $L^{\rm{gauge+fields}}_{\infty}$ structure the theory can be cast in an $L_{3}$ algebra, where new brackets related to the generalized structure constants and the double Lorentz transformations are computed. When one also considers the equation of motion of the fields the algebraic structure is exactly promoted to an $L_{4}$ algebra. 

This work is organized as follows: in Section \ref{GKSAA} we introduce GDFT in the generalized metric/flux formulation. Here we present the generalized Kerr-Schild ansatz (GKSA) for flat backgrounds. In Section \ref{Linf} we start by reviewing the way to obtain the products for a generic $L_{\infty}$ algebra. Then we cast the algebraic structure of both DFT and GDFT when the GKSA is considered. The present computations show the algebraic structure of the fundamental charged heterotic string and (Bosonic) Enhanced Double Field Theory, as we show in Section \ref{App}. Finally in Section \ref{Con} we summarize our work.

\section{The generalized Kerr-Schild ansatz}
\label{GKSAA}
\subsection{The DFT approach in metric formalism}
The GKSA is given by an exact and linear perturbation of the generalized background metric $H_{M N}$ ($M,N=0, \dots, 2D-1$) and an exact perturbation of the generalized background dilaton $d_{o}$. In this work we will consider linear perturbations in both fields. 

The perturbation of the generalized background metric $H_{M N}$ is given by a pair of generalized vectors, $K_{M}$ and $\bar{K}_{M}$, and an order parameter $\kappa$, such that
\bea
{\cal H}_{MN} = H_{MN} + \kappa (\bar{K}_{M} K_{N} + {K}_{M} \bar{K}_{N} ) \, ,
\label{DFTKS}
\eea
while the vectors satisfy 
\bea
\bar{K}_{M} & = & \frac12({\eta}_{M N} + {H}_{M N}) \bar{K}^{N} = \bar{P}_{M N} \bar{K}^{N} \, , \nn \\ 
K_{M} & = & \frac12({\eta}_{M N} - {H}_{M N}) K^{N} = {P}_{M N} {K}_{N} \, ,
\eea
and the generlized null conditions,
\bea
\eta^{MN} \bar{K}_{M} \bar{K}_{N} & = & \eta^{MN} K_{M} K_{N} =  \eta^{MN} \bar{K}_{M} K_{N} = 0 \, .
\label{nulldft}
\eea 
or, equivalently,
\bea
H^{MN} \bar{K}_{M} \bar{K}_{N} & = & H^{MN} K_{M} K_{N} =  H^{MN} \bar{K}_{M} K_{N} = 0 \, .
\label{nullHdft}
\eea
The generalized background dilaton $d_{o}$ is perturbed in a similar way\footnote{We work with a linear perturbation for simplicity, \textit{i.e.,} $f=\textrm{const.}$. In the general case $f = \sum_{n=0}^{\infty}\kappa^{n}f_{n}$},
\be
d = d_{o} + \kappa f\, .
\label{dilaton}
\ee

The perturbations of the GKSA satisfy the following extra conditions
\bea
{\bar K}^{P} \partial_{P} K^{M} + K_{P} \partial^{M}{\bar K}^{P} - K^{P} \partial_{P}{\bar K}^{M} & = & 0 \nn \, , \\
K^{M} \partial_{M}f = {\bar K}^{M} \partial_{M}f & = & 0 \, ,
\label{geodesic2}
\eea
which play the role of generalized geodesic equations.

In addition to the global $O(D,D)$ symmetry, the principle action of DFT is invariant under generalized diffeomorphisms generated infinitesimally  by $\xi^{M}$  through the generalized Lie derivative. Acting on an arbitrary vector it reads,
\bea
{\cal L}_\xi V_M = \xi^{N} \partial_N V_M + (\partial_M \xi^N - \partial^N \xi_{M}) V_N + \omega (\partial_{N} \xi^{N})V_{M} \, ,
\label{glie}
\eea 
where $\omega$ is a weight constant. The generalized metric ${\cal H}_{M N}$ and the generalized background metric $H_{M N}$ are tensors with $\omega=0$ with respect to generalized diffeomorphisms, and $\omega(e^{-2d})= \omega(e^{-2d_{o}})=1$. It is straightforward to check that conditions (\ref{geodesic2}) are covariant under generalized diffeomorphism transformations.

The Lagrangian of DFT is defined as,
\bea
{\cal L}_{DFT} & = & e^{-2d}(\frac18 {\cal H}^{MN} \partial_{M}{\cal H}^{KL}\partial_{N}{\cal H}_{KL} - \frac12 {\cal H}^{MN}\partial_{N}{\cal H}^{KL}\partial_{L}{\cal H}_{MK} \nn \\ && + 4 {\cal H}^{MN} \partial_{M}d\partial_{N}d  - 2 \partial_{M}{\cal H}^{MN} \partial_{N}d)  \, ,
\label{scalarDFT}
\eea
while the equations of motion can be written in terms of generalized curvatures,
\bea
{\cal R}_{\un P \ov Q} = {\cal P}_{P}{}^{M} \bar {\cal P}_{Q}{}^{N} \Big(\frac 18 \partial_M {\cal H}^{KL} \partial_N {\cal H}_{KL} -
\frac 1 4 (\partial_L - 2 (\partial_L d)) ({\cal H}^{LK} \partial_K {\cal
H}_{MN}) + 2 \partial_M \partial_N d \nn \\
- \frac 1 2 \partial_{(M|} {\cal H}^{KL} \partial_L {\cal H}_{|N)K}  + \frac
1 2 (\partial_L - 2 (\partial_L d)) ({\cal H}^{KL} \partial_{(M} {\cal H}_{N)K}
+ {\cal H}^K{}_{(M|}\partial_K {\cal H}^L{}_{|N)})\Big) = 0 \, ,
\label{Ricci}
\eea
and
\bea
{\cal R} = && \frac18 {\cal H}^{MN} \partial_{M}{\cal H}^{KL}\partial_{N}{\cal H}_{KL} - \frac12 {\cal H}^{MN}\partial_{N}{\cal H}^{KL}\partial_{L}{\cal H}_{MK} + 4 {\cal H}^{MN} \partial_{M}\partial_{N}d \nn \\ && + 4 \partial_{M}{\cal H}^{MN} \partial_{N}d - 4 {\cal H}^{MN} \partial_{M}d \partial_{N}d -  \partial_{M} \partial_{N} {\cal H}^{MN} = 0 \, .
\label{scalar}
\eea

\subsection{Extension to GDFT in flux formalism}

The ansatz (\ref{DFTKS}) and (\ref{dilaton}) are powerful tools to work pertubatively since the generalized null and geodesic conditions provide finite contributions to the action principle and the equations of motion. Interestingly enough (\ref{DFTKS}) admits an extension to the flux formulation of DFT, which is a mandatory step to consider a GDFT. In this case we consider perturbations of the form,
\bea
{\cal E}_{M}{}^{\ov A} = E_{M}{}^{\ov A} + \frac12 \kappa E_{M}{}^{\underline B} K_{\underline B} {\bar K}^{\ov A} \, , \nn \\ {\cal E}_{M}{}^{\underline A} = E_{M}{}^{\underline A} - \frac12 \kappa E_{M}{}^{\ov B} {\bar K}_{\ov B} K^{\underline A}\, ,
\label{GKSA}
\eea
where $K_{\un A} = {\cal E}^{M}{}_{\underline A} K_{M}={E}^{M}{}_{\underline A} K_{M}$ and $\bar{K}_{\ov A} = {\cal E}^{M}{}_{\ov A} \bar{K}_{M}=E^{M}{}_{\ov A} \bar{K}_{M}$ and ${\cal E}_{MA}$ is an $O(D,D+n)/O(D-1,1)_{L} \times O(1,D-1+n)_{R}$ frame. Here $\underline A= 0, \dots, D-1$ and $\ov A=0, \dots, D-1+n$ are $O(D-1,1)_{L}$ and $O(1,D-1+n)_{R}$ indices, respectively. In agreement with the previous section, we are going to consider a constant generalized frame background, \textit{i.e.}, $\partial_{M}E_{N A}=0$ and a constant generalized dilaton background $\partial_{M}{d_{o}}=0$ . 

Defining $\eta_{A B}$ and $H_{A B}$ as the invariant metrics of $O(D-1,1)_{L} \times O(1,D-1+n)_{R}$, we have, 
\bea
\eta_{AB} = {\cal E}_{M A}\eta^{MN} {\cal E}_{N B} = E_{M A}\eta^{MN} E_{N B} \, , \\
{H}_{AB} = {\cal E}_{MA} {\cal H}^{MN} {\cal E}_{N B} = E_{MA} {H}^{MN}E_{N B} \, .
\eea

The generalized fluxes take the form
\bea
{\cal F}_{ABC} & = & 3 {\cal E}_{[A}({\cal E}^{M}{}_{B}){\cal E}_{M C]} + \sqrt{2} f_{M N P} {\cal E}^{M}{}_{A} {\cal E}^{N}{}_{B} {\cal E}^{P}{}_{C} \, , \nn \\ {\cal F}_{A} & = & \sqrt{2}e^{2d}\partial_{M}\left({\cal E}^{M}{}_{A}e^{-2d}\right)  \, ,
\eea
where $f_{M N P}$ plays the role of generalized structure constants and therefore satisfy
\bea
f_{MNP}=f_{[MNP]}\, , \qquad f_{[MN}{}^{ R}f_{{P}]R}{}^{Q}=0\, , \label{consf}
\eea
and 
\bea
f_{ {MN}}{}^{ P}\partial_{P}\cdots =0 \, .
\label{fcond}
\eea
The generalized Lie derivative is deformed as,
\bea
\widehat{\cal L}_\xi V_M = {\cal L}_{\xi} V_{M} + f_{M N P} \xi^{N} V^{P} \, ,
\label{glie2}
\eea 
and, in addition, the theory is invariant under $O(D-1,1)_{L} \times O(1,D-1+n)_{R}$ or double Lorentz transformations,
\bea
\delta_{\Gamma} V^{A} = V^{B} \Gamma_{B}{}^{A} \, ,
\eea
where $V^{A}$ is a generic vector and $\Gamma_{A B} = - \Gamma_{B A}$ an arbitrary parameter. The previous transformations close with the following parameters 
\bea\label{par0}
\xi^{ M}_{12} & = & [\xi_1, \xi_{2} ]^{ M}_{(C_f)} \, ,\\
\Gamma_{12 { A} { B}} & = & 2 \xi_{[1}^{ P} \partial_{ P} \Gamma_{2] { A} { B}} - 2 \Gamma_{[1  A}{}^{ C} \Gamma_{2] {C B}}
 \, , \label{Lorentzbrack}
 \eea
where the $C_f$-bracket is a deformation of the $C$-bracket given by
\bea
 [\xi_1, \xi_{2} ]^{ M}_{(C_{f})}=2\xi^{ P}_{[1}\partial_{ P}\xi_{2]}^{ M}-\xi_{[1}^{ N}\partial^{ M}\xi_{2] N}+f_{{ PQ}}{}^{ M} \xi_{1}^{ P} \xi_2^{ Q}\, , \label{Cbrack}
\eea
where (\ref{fcond}) is required for consistency.

A flat covariant derivative acting on a generic vector is given by
\bea
{\cal D}_{A} V_{B} = {\cal E}_{A} V_{B} + {\cal W}_{AB}{}^{C} V_{C} \, , 
\label{covder}
\eea
where ${\cal E}_{A} = \sqrt{2} {\cal E}^{M}{}_{A} \partial_{M}$. The covariant derivative as well as the flat derivative can also be defined for background fields in a similar fashion. In (\ref{covder}) ${\cal W}_{AB}{}^{C}$ is the generalized spin connection, which is partially identified with the generalized fluxes according to
\bea
{\cal W}_{[ABC]} & = & -\frac13{\cal F}_{ABC}\, , \\
{\cal W}_{BA}{}^{B} & = &  - {\cal F}_{A}\, ,
\eea
in order to have, on the one hand, frame compatibility under covariant derivation and, on the other, partial integration with respect to the dilaton density, \textit{i.e.},
\be
\int e^{-2d}V {\cal D}_{ A}V^{ A}=-\int  e^{-2d}V^{ A} {\cal D}_{ A}V\, .
\ee
Considering the flat projectors as $P_{A B}=\frac12 \eta_{A B} - \frac12 H_{A B}$ and $\bar P_{A B}=\frac12 \eta_{A B} + \frac12 H_{A B}$, and the notation 
\bea
V_{A}=V_{\un A} + V_{\ov A}= P_{\un A}{}^{\un B} V_{\un B} + \bar P_{\ov A}{}^{\ov B} V_{\ov B} \, ,
\eea
the generalized curvatures (\ref{Ricci}) and (\ref{scalar}) are rewritten as
\bea
\label{GRicci_scalar}
{\cal R} & = & 2{\cal E}_{\underline{A}}{\cal F}^{\underline{A}} + {\cal F}_{\underline{A}}{\cal F}^{\underline{A}} - \frac16 {\cal F}_{\underline{ABC}} {\cal F}^{\underline{ABC}} - \frac12{\cal F}_{\ov{A}\underline{BC}}{\cal F}^{\ov{A}\underline{BC}} \, , \\
{\cal R}_{\ov{A}\un{B}} & = & {\cal E}_{\ov{A}}{\cal F}_{\un{B}} - {\cal E}_{\un{C}}{\cal F}_{\ov{A}\un{B}}{}^{\un{C}} + {\cal F}_{\un{C}\ov{DA}}{\cal F}^{\ov{D}}{}_{\un{B}}{}^{\un{C}} - {\cal F}_{\un{C}}{\cal F}_{\ov{A}\un{B}}{}^{\un{C}} \, .
\label{GRicci_tensor}
\eea
while the relevant projections of the fluxes are written in terms of $K_{A}$ and $\bar K_{\bar A}$ in the following way,
{\footnotesize
\bea
{\cal F}_{\un{ABC}} & = & \sqrt{2} f_{M N P} \Bigg( { E}^{M}{}_{{\un A}} { E}^{N}{}_{{\un B}} {E}^{P}{}_{\un C} -\frac{1}{2} \kappa K_{{\un A}} {\ov K}_{{\ov B}} E^{M{\ov B}} E^{N}{}_{{\un B}} E^{P}{}_{{\un C}} -\kappa   E^{M}{}_{{\un A}}K_{[{\un B}|} {\ov K}_{{\ov C}} E^{N{\ov C}} E^{P}{}_{|{\un C}]} \Bigg), \nn \\
{\cal F}_{\un{A}\ov{BC}} & = & \kappa\left(\bar K{}_{[\ov{C}}D{}_{\ov{B}]}K_{\un{A}} + K_{\un{A}}E_{[\ov{B}}\bar K_{\ov{C}]} \right) + \sqrt{2} f_{M N P} \Bigg( { E}^{M}{}_{{\un A}} { E}^{N}{}_{{\ov B}} {E}^{P}{}_{\ov C} - \frac12 \kappa K_{\un A} \bar K_{\ov D} E^{M \ov D} E^{N}{}_{\ov B} E^{P}{}_{\ov C} \Bigg)\, , \nn \\
{\cal F}_{\ov{A}\un{BC}} & = &  - \kappa\left(K_{[\un{C}}D_{\un{B}]}\bar K_{\ov{A}} + \bar K_{\ov{A}}E_{[\un{B}}K_{\un{C}]} \right) + \sqrt{2} f_{M N P} \Bigg( { E}^{M}{}_{{\ov A}} { E}^{N}{}_{{\un B}} {E}^{P}{}_{\un C} + \frac12 \kappa \bar K_{\ov A} K_{\un D} E^{M \un D} E^{N}{}_{\un B} E^{P}{}_{\un C} \Bigg) \, , \nn \\
{\cal F}^{\un{A}} & = &  - \frac{1}{2}\kappa\left( (E_{\bar B}\bar K^{\ov B})K^{\un A} + (E_{\ov B} K^{\un A}) \bar K^{\ov B} + 4E^{\un{A}}f\right)\, .
\label{constrained_fluxes}
\eea}
The flat version of the null conditions reads 
\bea
K_{\un A} K^{\un A} & = & \bar K_{\ov A} \bar K^{\ov A} = 0 \, , \label{flatnull}   
\eea
and the flat geodesic conditions now contain a contribution related to the generalized structure constants,
\bea
 K^{\underline A} E_{\underline A} {\bar K}^{\overline C} + \sqrt{2} {K}^{\un A} \bar K^{\ov B} f_{M P Q} E^{M}{}_{\un A} E^{P}{}_{\ov B} E^{Q \ov C}  & = & 0 \, , \\ 
{\bar K}^{\overline A} E_{\overline A} {K}^{\underline C} + \sqrt{2} {\bar K}^{\ov A}  K^{\un B} f_{M P Q} E^{M}{}_{\ov A} E^{P}{}_{\un B} E^{Q \un C}  & = & 0 \, , \\
K^{\underline A} E_{\underline A} f = {\bar K}^{\ov A} E_{\overline A} f & = & 0 \, .
\label{flatgeo}
\eea

\section{$L_{\infty}$ algebras}
\label{Linf}
In this section we start by reviewing how to fit DFT in an $L_{\infty}$ algebra and then we show the extension to GDFT. We always consider the GKSA in order to obtain closed expressions when dynamics is taken into account and we dedicate next section to discuss about the family of theories that can be described within this approach. 

\subsection{Basics}
\label{Basics}
Let us consider a vector graded space $X$ which is the direct sum of vector spaces $X_n$, each of which has degree $n$
\be
X = \bigoplus_{n}  X_n \,, \quad  n \in \mathbb{Z} \ .
\ee
We will denote by $x$ an element of $X$ with definite degree, $i.e$, $x\in X_p$ for some fixed $p$.
We consider multilinear products $\ell_k$
\be
\ell_k:  X^{\otimes k} \rightarrow X \ ,
\ee
with degree given by
\be
\hbox{deg}(\ell_k(x_1,x_2,...,x_k))= k-2 + \sum_{i=1}^k  \hbox{deg} (x_i) \ .
\ee
For a permutation $\sigma$ of $k$ labels we have
\be
\ell_k ( x_{\sigma(1)} , \ldots , x_{\sigma(k)} ) \ = \ (-1)^\sigma \epsilon(\sigma;x ) \,
\ell_k (x_1 , \ldots \,, x_k) \ .
\ee
The $(-1)^\sigma$ factor gives a plus or minus sign if the permutation is even or odd, respectively. The $\epsilon(\sigma;x )$ factor is the Koszul sign. For a graded commutative algebra $\Lambda (x_1, x_2, \cdots )$ with
\be
x_i \wedge x_j \ = \ (-1)^{{\rm deg}(x_i) {\rm deg}(x_j)}   \,  x_j \wedge x_i \,,   \quad \forall i, j\  ,
\ee
the Koszul sign for a general permutation is given by
\be
 x_1\wedge \ldots  \wedge x_k   =  \epsilon (\sigma; x)  \   x_{\sigma(1)} \wedge \ldots   \wedge \, x_{\sigma(k)} \ .
\ee
It is convenient to abuse with the notation in the following way
\be
\ (-1)^{{\rm deg}(x_i) {\rm deg}(x_j)}\equiv(-1)^{x_i x_j}\ .
\ee

The $L_\infty$ relations are labeled by a positive integer $n$ given by the number of inputs. Explicitly they are
\be
\label{main-Linty-identity}
\sum_{i+j= n+1}  (-1)^{i(j-1)} \sum_\sigma  (-1)^\sigma \epsilon (\sigma; x) \, \ell_j \, \bigl( \, \ell_i ( x_{\sigma(1)}  \,, \, \ldots\,, x_{\sigma(i)} ) \,, \, x_{\sigma(i+1)}, \, \ldots \, x_{\sigma (n)} \bigr) \ = \ 0\ .
\ee
The sum over $\sigma$ is a sum over ``unshuffles'', it includes only the terms which satisfy
\be
\sigma(1) <  \, \cdots \, <  \, \sigma(i) \,,  \qquad
\sigma(i+1) <  \, \cdots \, <  \, \sigma(n) \ .
\ee
It is common to write these relations as
\be
\label{main-Linty-identity-schem}
\sum_{i+j= n+1}  (-1)^{i(j-1)}  \ell_j \, \ell_i \ = \ 0\ ,
\ee
such that
\bea
n = 1 \ \ \ \ \ \ \ \ 0 &=& \ell_1 \ell_1 \\
n = 2 \ \ \ \ \ \ \ \ 0 &=& \ell_1 \ell_2 - \ell_2 \ell_1 \\
n= 3 \ \ \ \ \ \ \ \  0 &=& \ell_1 \ell_3 + \ell_2 \ell_2 + \ell_3 \ell_1 \\
n = 4 \ \ \ \ \ \ \ \ 0 &=& \ell_1 \ell_4 - \ell_2 \ell_3 + \ell_3 \ell_2 - \ell_4 \ell_1 \ , \ \dots
\eea

For instance, the $n=3$ case is given by
 \bea
  0  & = & \ell_2(\ell_2(x_1,x_2),x_3) + (-1)^{(x_1+ x_2) x_3}\ell_2(\ell_2(x_3,x_1),x_2)
   +(-1)^{(x_2+ x_3) x_1 }\ell_2(\ell_2(x_2,x_3),x_1) \nn \label{L3L1}\\
   &&\!\!\!\!\!\!\!\!\!\!\!\!\!\!\!\!\!\!\!\! +  \ell_1(\ell_3 (x_1,x_2, x_3))    + \ell_3(\ell_1 (x_1) ,x_2, x_3)
  +  (-1)^{x_1}  \ell_3( x_1 ,\ell_1(x_2), x_3)
    +  (-1)^{x_1+ x_2}  \ell_3( x_1 ,x_2, \ell_1(x_3))  \ . \nn
\label{n=3}
 \eea

One must assign a given degree $p$ to gauge parameters, fields, EOM's, etc., and so specify to what vector subspace $X_p$ they belong. In this work we consider that the space of degree two contains the constants ($c$), the space of degree one contains
functions ($\chi$), the space of degree zero contains the gauge parameters ($\zeta$), the space of degree minus one contains the fields ($\Psi$) and, finally, the space of degree minus two the dynamics (${\cal F}$).  In general the products can be read from the symmetries and dynamics of a given field theory. The symmetry transformations define the brackets $\ell_{n+1}(\z,  \Psi^n)$ as follows
\be
        \delta_{\xi}\Psi =\sum_{n\ge 0}   \frac{1}{n!}
      (-1)^{{n(n-1)}/{2}}\,			
			 \ell_{n+1}(\xi,  \Psi^n)
			  \ , \label{gaugel}
\ee
where $
\Psi^k = \underbrace{\Psi,...,\Psi}_{k\;\text{times}}$. The equations of motion define the $l_n (\Psi^n)$ brackets as follows
\be
  {\cal F}(\Psi)  =  \sum_{n=1}^\infty
   \frac{(-1)^{n(n-1)/ 2}}{n!}  \ell_n(\Psi^n) \, .
   \label{eoml}
  \ee
Both (\ref{gaugel}) and (\ref{eoml}) are fundamental relations that can be used to read non-trivial products, and then extra products can appeared upon checking the $L_{\infty}$ relations (\ref{main-Linty-identity}).

\subsection{GKSA-DFT as an $L_{3}$ algebra}
\label{GKSA-DFT}
Here we follow the construction presented in \cite{Hohm:2017pnh}. In that work the authors show that when the arguments of $l_2$ are the DFT gauge parameters, this product is related to the C-bracket. Moreover, the first line in (\ref{n=3}) coincides with the Jacobiator and the last line characterizes the  non-trivial Jacobiator of DFT given by
\be
J(\xi_{1},\xi_{2},\xi_3)^{M} = \frac{3}{2}\partial^{M}(\xi_{[1}^{N} \xi_{2}^{P} \partial_{N} \xi_{3]P}) = N^M \, .
\ee
Considering the following relation derived from (\ref{gaugel}),(\ref{eoml}) and (\ref{main-Linty-identity}),
\bea
\label{commurel}
[\delta_{\zeta_1},\delta_{\zeta_2}] \Psi=\delta_{-\mathbf C(\zeta_1,\zeta_2)} \Psi \, ,
\eea
with $\mathbf C(\zeta_1,\zeta_2) \equiv \;\ell_{2}(\zeta_1,\zeta_2)\ ,$
the non-trivial products are
\bea
\ell_{1}(\chi) = \partial \chi \in X_{0}, \\
\ell_{1}(c) = \iota c \in X_{1},\\
\ell_{2}(\xi_{1},\xi_{2}) = \big[ \xi_{1},\xi_{2} \big]_{C} \in X_{0},\\
\ell_{2}(\xi , \chi) = \frac{1}{2} \xi^{K}\partial_{K}\chi \in X_{1},\\
\ell_{3}(\xi_{1},\xi_{2},\xi_{3}) = - N(\xi_{1},\xi_{2},\xi_{3}) \in X_{1}.
\eea

On the other hand, considering
\bea
\delta_{\xi} {\cal H}_{M N} & = & \xi^{P} \partial_{P} {\cal H}_{M N} + 2 (\partial_{(M} \xi^{P} - \partial^{P} \xi_{(M}) {\cal H}_{N) P} \nn \\
\delta_{\xi} d & = & \xi^{P} \partial_{P}d - \frac12 \partial_{P}\xi^{P} \, , 
\eea
and (\ref{gaugel}), and invoking the GKSA it is straightforward to find
\bea
\ell_{1}(\xi)_{t} & = & 2(\partial_{\un M} \xi_{\ov N} - \partial_{\ov N} \xi_{\un M}) \, , \\ 
\ell_{1}(\xi)_{s} & = & - \frac12 \partial_{P} \xi^{P} \, ,  \\
\ell_{2}(\xi,K)_{v} & = & \delta_{\xi} K_{M} \, , \\
\ell_{2}(\xi,\bar K)_{v} & = & \delta_{\xi} \bar K_{M} \, , \\
\ell_{2}(\xi,f)_{s} & = & \xi^{P}\partial_{P}f \, ,
\eea
where the letters $s$,$v$,$t$ means that we are considering the scalar, vectorial or tensorial part of the product, respectively. 

\subsubsection{Pertubative DFT as an exact $L_{3}$ algebra}

The closed expressions for the dynamics can be easily obtained from (\ref{eoml}). Considering the equation of motion for the generalized dilaton we identify, 
\bea
\ell_{1}(f)_s & = & 4\kappa{H}^{K L} \partial_{K}\partial_{L}{f} \\
\ell_{2}(f,f)_s & = & 8\kappa^{2} {H}^{K L} \partial_{K}{f} \partial_{L}{f} \\
\ell_{2}(\bar K,K)_s & = & 4 \kappa \partial_{K} \partial_{L} (  K^{K} \bar{K}^{L})  \, ,
\eea
and, analogously, from the generalized metric equation we obtain,
\bea
\ell_{1}(f)_t & = & 4 \kappa P_{K}{}^{M} \bar{P}_{L}{}^{N}\partial_{M N}{f}  \\
\ell_{2}(\bar K,K)_t & = & \kappa \Big[ {H}^{M N} \partial_{M N}\big(K_{K} \bar{K}_{L}\big)-2\partial_{M N} \big( K^{N} \bar{K}_{L} P_{K}{}^{M} - K_{K} \bar{K}^{N} \bar{P}_{L}{}^{M}\big) \Big]  \\
\ell_{3}(f,\bar K, K)_{t} & = & -6 \kappa^{2} \Big[\ {H}^{MN}\partial_{M}f \partial_{N}\big( K_{K} \bar{K}_{L}\big) -2 P_{K}{}^{M} \partial_{M}\big( K^{N}\bar{K}_{L}\partial_{N}f\big) \nn \\ && + 2\bar{P}_{L}{}^{M}\partial_{M}\big( K_{K}\bar{K}^{N}\partial_{N}f\big) \ \Big]  \, ,
\eea
where $\partial_{M N}= \partial_{M} \partial_{N}$.

In order to verify the $L_{\infty}$ relations given by (\ref{main-Linty-identity})
it is necessary to only include extra products related to the gauge transformation of the equations of motion, $\ell_{2}(\xi,{\cal R})_s=\delta_{\xi}{\cal R}$ and $\ell_{2}(\xi,{\cal R}_{\un M \ov N})_t=\delta_{\xi}{\cal R}_{\un M \ov N}$, while the remaining products are null.

\subsection{GKSA-GDFT as an $L_{4}$ algebra}
\label{GKSA-GDFT}
The extension to GDFT will be performed in several steps. We start by considering only the subspaces related to the brackets algebra ($X_2$,$X_1$,$X_0$), then we include the subspace of fields $X_{-1}$ and finally we include both fields and their dynamics $X_{-2}$.

\subsubsection{GDFT bracket algebra as an $L_{3}$ algebra}
We start by discussing the
subalgebra corresponding to the pure gauge structure, given by the $C_f$-bracket algebra (\ref{Cbrack}) and the Double Lorentz bracket (\ref{Lorentzbrack}). The graded vector space is taken to contain three spaces of fixed degree,
\bea
0 & \rightarrow X_2 & \rightarrow X_1 \rightarrow X_0 \nn \\ & \quad c & \quad \, \, \, \chi \, \, \, \quad \quad \, \zeta
\eea
where $\zeta=(\xi,\Lambda)$ is a generic parameter and the above arrows define the $\ell_1$ action.  From $X_2$ to $X_1$ the action is given by the inclusion map, while from $X_1$ to $X_0$ the action is given by the partial derivative. Acting on $X_0$ the map $\ell_1$ is null since we are not considering the fields yet. At this level the non-trivial products are
\bea \label{prod1}
\ell_{1}(\chi) = \partial \chi \in X_{0}, \\
\ell_{1}(c) = \iota c \in X_{1},\\
\ell_{2}(\xi_{1},\xi_{2}) = \big[ \xi_{1},\xi_{2} \big]_{C_{f}} \in X_{0},\\
\ell_{2}(\xi , \chi) = \frac{1}{2} \xi^{K}\partial_{K}\chi \in X_{1},\\
\ell_{3}(\xi_{1},\xi_{2},\xi_{3}) = - N_{o}(\xi_{1},\xi_{2},\xi_{3}) - N_{f}(\xi_{1},\xi_{2},\xi_{3}) \in X_{1} \, , \\
\ell_{2}(\xi,\Gamma) =  \xi^{ P} \partial_{ P} \Gamma_{{ A} { B}} \in X_{0}, \label{extra}\\
\ell_{2}(\Gamma_{1},\Gamma_{2}) = - \Gamma_{1  A}{}^{ C} \Gamma_{2 {C B}} \in X_{0} \, , \label{prod7}
\eea
where $N_o$ and $N_f$ can be computed from the Jacobiator of GDFT,
\be
J(\xi_{1},\xi_{2},\xi_3)^{M} = \frac{3}{2}\partial^{M}(\xi_{[1}^{N} \xi_{2}^{P} \partial_{N} \xi_{3]P} + \frac13 f_{N P Q} \xi_{1}^{N} \xi_{2}^{P} \xi_{3}^{Q}) = N_{o}^M + N_{f}^M \, .
\ee

The bracket $\Gamma_{[1A}{}^{C} \Gamma_{2]C B}$ encodes the algebra of matrix multiplication and therefore the analogous of the Jacobiator for the Double Lorentz symmetry is trivially null. Moreover from (\ref{extra}) it is straightforward to show the following relation, 
\bea
\ell_{2}(\Gamma,\partial \chi)=\partial \ell_{2}(\Gamma,\chi) \, . 
\eea
Using the previous relation and the products (\ref{prod1})-(\ref{prod7}) it is straightforward to show that the relations $n\geq4$ are trivial.

\subsubsection{Off-shell GDFT as extended $L_{3}$ algebra}
Now we extend the $L_{3}$ algebra describing the $C_{f}$ and the double Lorentz brackets to include the fields and the symmetry
transformations. We recall at this point that the generalized metric formalism has to be abandoned in order to describe the GDFT structure.  The graded vector space now  contains four spaces,
\bea
0 \rightarrow & X_2 & \rightarrow X_1 \rightarrow X_0 \rightarrow X_{-1} \nn \\ & c & \quad \, \, \, \chi \, \, \, \quad \quad \zeta \, \, \, \quad \, \, \, \Psi
\eea
where $\Psi=(K,\bar K , f)$ and $\ell_{n}\Psi^{n}=0$ with $n\geq1$ since there is no dynamics at this point. From the symmetry transformations we read the following products,
\bea
\ell_{1}(\xi)_{t} & = & 2(\partial_{\un M} \xi_{\ov N} - \partial_{\ov N} \xi_{\un M}) \, , \\ 
\ell_{1}(\xi)_{s} & = & - \frac12 \partial_{P} \xi^{P} \, , \\
\ell_{2}(\xi,\bar K)_{\ov v} & = &  \widehat{\mathcal L}_{\xi} \bar K_{\ov A}  \, , \\
\ell_{2}(\Gamma,\bar K)_{\ov v} & = &  \delta_{\Gamma} \bar K_{\ov A} \, , \\
\ell_{2}(\xi, K)_{\un v} & = & \widehat{\mathcal L}_{\xi} K_{\un A} \, , \\
\ell_{2}(\Gamma, K)_{\un v} & = & \delta_{\Gamma} K_{\un A} \, ,  \\
\ell_{2}(\xi,f)_{s} & = & \xi^{P}\partial_{P}f \, .
\eea
The $L_{\infty}$ relations can be probed considering the previous list and the one from the previous section. The relations $n=1$ and $n=2$ are trivial. The relation $n=3$ is not trivial for the case $x_1=\Psi$, $x_2=\zeta_2$, $x_3=\zeta_3$,
 \bea
  0  & = & \ell_2(\ell_2(\Psi,\zeta_2),\zeta_3) + \ell_2(\ell_2(\zeta_3,\Psi,),\zeta_2)
   +\ell_2(\ell_2(\zeta_2,\zeta_3),\Psi)  
    \ . \nn
 \eea
The previous expression can be rewritten in the following form,
\bea
[\delta_{\zeta_2},\delta_{\zeta_3}]\Psi = \delta_{\zeta_{23}} \Psi
\eea
and therefore it is satisfied using the closure condition for the deformed generalized diffeomorphisms and double Lorentz transformations. Relations with $n \geq 4$ are trivial.

\subsubsection{Pertubative GDFT as an exact $L_{4}$ algebra}
Finally we extend the $L_{3}$ algebra describing the $C_{f}$ and the double Lorentz brackets algebra to include the dynamics. The graded vector space now  contains five spaces,
\bea
0 \rightarrow & X_2 & \rightarrow X_1 \rightarrow X_0 \rightarrow X_{-1} \rightarrow X_{-2} \nn \\ & c & \quad \, \, \, \chi \, \, \, \quad \quad \zeta \, \, \, \quad \, \, \, \Psi \quad \quad \, \, {\cal F}
\eea
where the perturbative equations of motion are related to the equations of the generalized dilaton and the generalized metric ${\cal F}=(\cal R, \cal R_{\ov A \un B})$, but considering the GKSA and the linear perturbation for the generalized dilaton. From (\ref{eoml}) we have
\bea
\ell_{1}(f)_s & = & - 4 \kappa  E^{\un A}\left(E_{\un{A}}f\right) \\ \ell_{2}(f,f)_s & = & - 8 \kappa^2 E_{\un{A}}f E^{\un{A}}f  \\  \ell_{2}(K,\bar K)_s & = & 2 \kappa  E^{\un A}\left(K_{\un A} E_{\ov B}\bar K^{\ov B} + \bar K^{\ov B} E_{\ov B} K_{\un A}\right) - 4 \kappa f_{\ov A \un B \un C} f^{\un D \un B  \un C} \bar K^{\ov A} K_{\un D} \\  \ell_{4}(K,K,\bar K,\bar K)_s & = &
- 6 \kappa^2 \bar K_{\ov B} \bar K_{\ov C} \left[ (E^{\ov C} K^{\un A}) E^{\ov B} K_{\un A}  - 2 f^{\ov B}{}_{\un B \un C} K_{{\un A}}  f^{\ov C \un A \un C} K^{{\un B}}  \right]
\eea
where we use the compact notation $f_{A B C} = E^{M}{}_{A} E^{N}{}_{B} E^{P}{}_{C} f_{M N P}$. The previous contributions come from the GDFT Lagrangian up to a cosmological term that requires a field redefinition. Analogously, from the generalized flat Ricci scalar we read
\bea
\ell_{1}(f)_t & = & 2\kappa \left[ \sqrt{2} f_{\ov A \un B \un C} E^{\un{C}}f - E_{\ov A}[E_{\un{B}}f] \right] \\ \ell_{2}(K,\bar K)_t & = & \kappa\left[ 2 f^{\ov D}{}_{\un B \un C}  K^{\un C} \bar K_{\ov E} f^{\ov E}{}_{\ov D \ov A} -2 f^{\un C}{}_{\ov D \ov A} \bar K^{\ov D} K_{\un D} f^{\un D}{}_{\un B \un C} \right. \nn \\ &&
- 2 \sqrt{2}  f^{\un C}{}_{\ov D \ov A} \left(- \frac12 K_{\un{C}}D_{\un{B}}\bar K^{\ov{D}} - \frac12 \bar K^{\ov{D}}E_{\un{B}}K_{\un{C}} + \frac12 K_{\un{B}}D_{\un{C}}\bar K^{\ov{D}} \right. \nn \\&& \left. + \frac12 \bar K^{\ov{D}}E_{\un{C}}K_{\un{B}} \right) - \sqrt{2} \left((E_{\bar B}\bar K^{\ov B})K^{\un C} + (E_{\ov B} K^{\un C}) \bar K^{\ov B} \right)  f_{\ov A \un B \un C} \nn \\ && +2 E^{\un C}\left[ - K_{[\un{C}}D_{\un{B}]}\bar K_{\ov{A}} - \bar K_{\ov{A}}E_{[\un{B}}K_{\un{C}]} + \frac{1}{\sqrt{2}} \bar K_{\ov A} K_{\un D} f^{\un D}{}_{\un B \un C} \right] \nn \\
&&
+ E_{\ov A}\left[(E_{\bar C}\bar K^{\ov C})K_{\un B} + (E_{\ov C} K_{\un B}) \bar K^{\ov C} \right]
\nn \\&& \left. - 2 \sqrt{2} f^{\ov D}{}_{\un B \un C}\left(\bar K{}_{[\ov{A}}D{}_{\ov{D}]}K^{\un{C}} + K^{\un{C}}E_{[\ov{D}}\bar K_{\ov{A}]}  \right) \right] \\ \ell_{3}(f,K,\bar K)_t & = &  \kappa^2 \left[-\frac{3}{2} \left(4E^{\un{C}}f\right) \left( K_{\un{B}}D_{\un{C}}\ov{K}_{\ov{A}} \right) + 6 K_{\un C}\bar{K}_{\ov A} E^{\un C}\left[E_{\un{B}}f\right] \right. \nn \\ &&
- 6 \left( E^{\un{C}}f\right) K_{\un{B}}D_{\un{C}}\ov{K}_{\ov{A}} + 6 \left( E^{\un{C}}f\right) \ov{K}_{\ov{A}}E_{\un{B}}K_{\un{C}} \nn \\ && \left.
- 6 \left( E^{\un{C}}f\right) \ov{K}_{\ov{A}}E_{\un{C}}K_{\un{B}} - 6 \sqrt{2} \left( E^{\un{C}}f\right)  \bar K_{\ov A} K_{\un D} f^{\un D}{}_{\un B \un C} \right] \\ \ell_{4}(K,K, \bar K,\bar K)_t & = &  6 \kappa^2 \left[ \left( (E_{\ov B} K^{\un C}) \bar K^{\ov B} \right) \left( K_{\un{B}}D_{\un{C}}\ov{K}_{\ov{A}} \right)  -  K_{\un C}\bar{K}_{\ov A} E^{\un C}\left[(E_{\bar C}\bar K^{\ov C})K_{\un B} \right. \right. \nn \\ && \left. + (E_{\ov C} K_{\un B}) \bar K^{\ov C} \right]   -  K^{\un C} \bar K_{\ov D}  \bar K_{\ov{A}}E^{\ov D}\left[E_{\un{B}}K_{\un{C}}\right]
+  K^{\un C} \bar K_{\ov D} K_{\un{B}}E^{\ov D}\left[D_{\un{C}}\bar K_{\ov{A}}\right] \nn \\ &&
+ K^{\un C} \bar K_{\ov D} E^{\ov D} \left[\bar K_{\ov{A}}E_{\un{C}}K_{\un{B}} \right]
+ \sqrt{2} K^{\un C} \bar K_{\ov D} K_{\ov A} E^{\ov D} \bar  K_{\un D} f^{\un D}{}_{\un B \un C} \nn \\ &&
+  K_{\un{B}}D_{\un{C}}\bar K^{\ov{D}} \bar K{}_{\ov{A}}D{}_{\ov{D}}K^{\un{C}} + \bar K^{\ov{D}}E_{\un{C}}K_{\un{B}} K^{\un{C}}E_{\ov{D}}\bar K_{\ov{A}}  \nn \\
&&
+ \left( (E_{\ov B} K^{\un C}) \bar K^{\ov B} \right) K_{\un{B}}D_{\un{C}}\ov{K}_{\ov{A}} - \left((E_{\ov B} K^{\un C}) \bar K^{\ov B} \right) \ov{K}_{\ov{A}}E_{\un{B}}K_{\un{C}} \nn \\ &&
+ \left((E_{\bar B}\bar K^{\ov B})K^{\un C} + (E_{\ov B} K^{\un C}) \bar K^{\ov B} \right) \ov{K}_{\ov{A}}E_{\un{C}}K_{\un{B}}  \nn \\ && \left. + \sqrt{2} \left(\bar K^{\ov B} E_{\ov B} K^{\un C}  \right)  \bar K_{\ov A} K_{\un D} f^{\un D}{}_{\un B \un C} \, \right].
\eea
At this point we include products related to the gauge transformation of the equations of motion, $\ell_{2}(\xi,{\cal R})_s=\delta_{\xi}{\cal R}$, and $\ell_{2}(\xi,{\cal R}_{\ov A \un B})_t=\delta_{\xi}{\cal R}_{\ov A \un B}$, $\ell_{2}(\Lambda,{\cal R}_{\ov A \un B})_t=\delta_{\Lambda}{\cal R}_{\ov A \un B}$, where the last contribution can be easily computed considering that each index of ${\cal R}_{\ov A \un B}$ transforms as a projected double Lorentz vector. The products related to the transformation of the equations of motion are required to check the $n=2$ relation. In this context, the absence of a $\ell_{3}(\zeta_{1},\zeta_2,{\cal F})$ implies that the closure of the gauge algebra holds off-shell. The remaining products are also null as can be easily verified.

\section{Applications}
\label{App}
\subsection{Fundamental charged heterotic string}

The most simple theory that lies inside the family of low energy effective field theories that can be reproduced with the GKSA is the fundamental charged heterotic string solution in $D=10$ \cite{Sen}. The duality approach can be easily constructed considering the following parametrization,
\bea
{H}_{M N} = \left(\begin{matrix} g_{o}^{\mu \nu} & - g_{o}^{\mu \rho} C_{o\rho \nu} & -  g_{o}^{\mu \rho} A_{o \rho i} \\
- g_{o}^{\nu \rho} C_{o\rho \mu} &  g_{o\mu \nu} + C_{o\rho \mu} C_{o\sigma \nu}  g_{o}^{\rho \sigma} + A_{o\mu}{}^i \kappa_{ij} A_{o\nu}{}^j &
C_{o\rho \mu}  g_{o}^{\rho \sigma} A_{o\sigma i} + A_{o\mu}{}^j \kappa_{ji} \\
- g_{o}^{\nu \rho} A_{o\rho i} & C_{o\rho \nu} g_{o}^{\rho \sigma} A_{o\sigma i} +  A_{o\nu}{}^j \kappa_{ij} & \kappa_{ij} + A_{o\rho i} g_{o}^{\rho \sigma} A_{o\sigma j}\end{matrix}\right) \ ,
\label{Gmetric}
\eea
with $\kappa_{ij}$ a Cartan-Killing metric. Since the generalized structure constants force us to describe the theory with the generalized frame/flux formalism, compatibility with the ansatz forces  that the gauge field remains unperturbed as in \cite{KL}. Similarly, the generalized null vectors $K_{M}$ and ${\bar K}_{M}$ can be parametrized in terms of a pair of null vectors $l$ and $\bar l$ in the following way,
\bea
K_{M} = \, \frac{1}{\sqrt{2}} \left( \begin{matrix} l^{\mu} \\ - l_{\mu} - C_{o \rho \mu} l^{\rho } \\ - A_{oi \rho} {l}^{\rho} \end{matrix} \right) \, , \quad
\bar{K}_{M} = \, \frac{1}{\sqrt{2}} \left( \begin{matrix} {\bar l}^{\mu} \\  {\bar l}_{\mu} - C_{o \rho \mu} {\bar l}^{\rho} \\- A_{oi \rho} {\bar l}^{\rho} \end{matrix} \right) \, , 
\eea 
where $C_{o\mu \nu}=b_{o\mu \nu} + \frac12 A_{o\mu}{}^{i} A_{o\nu i}$ and the null vectors are constraint by (\ref{flatgeo}). The parametrization of the dilaton is $
 e^{-2d_{o}} = \sqrt{g_{o}} e^{-2 \phi}$. In this case the perturbed solution is given by
\bea
ds^{2} = \frac{1}{1+NH(r)}(-dt^{2}+ (dx^{9})^2) + \frac{q^2 H(r)}{4N(1+NH(r))^2}(dt+dx^{9})^2 + \sum_{i=1}^{8} dx^{i} dx^{i} \, , 
\eea
with $H(r)$ a Green function and $N$ a constant. The non-vanishing components of the two form and gauge field are
\bea
b_{9t} & = &\frac{NH(r)}{1+NH(r)} \,, \\
A^1_{0} & = & A^{1}_{9} = \frac{qH(r)}{1+NH(r)} \, ,
\eea
with $q$ a charge and $\phi=-\frac12 \textrm{ln}(1+N H(r))$. 

At the level of the symmetry transformations the algebraic structure of the duality covariant approach of this theory is an $L_{3}$ algebra, given by the transformations of $K_{M}$, $\bar K_{M}$ and $f$ under generalized diffeomorphisms and Double Lorentz transformations. While the former encodes ordinary diffeomorphisms and abelian/non abelian gauge transformations for $b_{0 \mu \nu}/A_{o\mu i}$, the latter transforms the flat version of the null vectors with a 10-dimensional Lorentz parameter $\Lambda_{ab}$ such that,
\be
\delta_{\Lambda} l_{a} =  \Lambda_{a}{}^{b} l_{b} \, ,  \quad \delta_{\Lambda} \bar l_{a} =  \Lambda_{a}{}^{b} \bar l_{b} \, ,
\ee
where $l_{a} = e_{o}^{\mu}{}_{a} l_{\mu}$ and $\bar l_{a} = e_{o}^{\mu}{}_{a} \bar l_{\mu}$. The full perturbative GDFT for this solution can be cast in an $L_{4}$ algebra with $\kappa^2$ contributions in the equations of motion as we showed in the previous section.

\subsection{(Bosonic) Enhanced DFT}
In compactifications of the bosonic string on k-dimensional tori,
the $U(1)_{L} \times U(1)_R$ symmetry of the Kaluza-Klein reduction gets enhanced at special points in moduli space and there are new massless scalars transforming in the adjoint representation of the enhanced symmetry groups. The new gauge group corresponds to $G_{L} \times G_{R}$ where $G_{L}=G_{R}$, and reduces to the standard $U(1)_{L} \times U(1)_R$ outside the points that provide the enhancement. As we mentioned, DFT incorpores T-duality
as a global symmetry group and therefore it is expected that there exists a formulation that captures these new states in a duality covariant way \cite{Enh}. When $G_{L}\times G_{R}$ has non-simple roots, it was shown in \cite{Mariana} that the C-bracket can be deformed in a way that preserves the duality covariance. The deformation accounts for
the cocycle factors that are necessary in the vertex representation of the current algebra. In terms of the generalized Lie derivative this deformation is
\bea
({\cal \widehat L}_{E_A}E_B)^M= ({\cal L}_{E_A}E_B)^M + \widehat \Omega_{AB}{}^C E^{M}{}_{C} \, ,
\eea
where $ \widehat\Omega_{ABC}$ vanishes if one or more indices correspond to Cartan generators and if $A,B,C$\footnote{We use the same notation, but these indices $A,B,C \dots$ must be thought as double internal index of a generalized parallelizable manifold.} are associated with roots of the enhancement algebra, say $\alpha,\beta,\gamma$, respectively,
\bea
\widehat\Omega_{ABC}=\left\{\begin{matrix} (-1)^{\alpha * \beta}\;\delta_{\alpha+\beta+\gamma} &\ \ {\rm if \ two \ roots \ are \ positive,}\\
-(-1)^{\alpha * \beta}\;\delta_{\alpha+\beta+\gamma} &\ \ {\rm if \ two \ roots \ are \ negative.} \\
  \end{matrix}\right.
\label{deform}
 \eea
 The tensor $\widehat\Omega_{ABC}$ satisfies
\bea
\widehat\Omega_{ABC}=\widehat\Omega_{[ABC]}\, , \qquad \widehat\Omega_{[AB}{}^D\widehat\Omega_{C]D}{}^E=0\, , \qquad \widehat\Omega_{ABC}\partial^C\cdots =0\, ,
\eea
and therefore (\ref{deform}) can be easily identified with the generalized structure constants $f_{ABC}$ upon trivially extended $O(D,D+n)\rightarrow O(D+n,D+n)$ in (\ref{glie2}). In this sense, the algebraic structure of enhanced DFT can be cast in the $L_{3}$ framework at the level of $C_{\Omega}$-bracket algebra according to the results of this work.

\section{Summary}
\label{Con}
In this work we show that GDFT can be cast in an $L_{\infty}$ structure. The presence of a deformed generalized Lie derivative and the double Lorentz transformation enrich the algebraic structure including non-trivial products to the well known $L_{\infty}$ structure of DFT. The frame formalism is needed to compute the generalized fluxes, which are deformed with a generalized version of the structure constants. At the level of the symmetry transformations the algebraic structure of GDFT is given by an $L_{3}$ algebra. We also show that the study of the dynamics can be performed in a closed form considering a GKSA for the generalized frame and a linear perturbation for the generalized dilaton. When dynamics is taken into account, the structure is promoted to an $L_{4}$ algebra with $\kappa^2$ corrections. The present computation has direct implications in the low energy effective action principle of the fundamental heterotic charged string, and bosonic string compactified on an specific internal k-dimensional torus with enhanced gauge symmetry. The latter is described by a enhanced DFT, which can be understood as a particular case of GDFT and therefore, at the level of the deformed bracket, the algebraic structure is an $L_{3}$ algebra.

\subsection*{Acknowledgements}
We thank D. Marqués for useful discussions. M.M. thanks the Max-Planck institute for kind hospitality during the final stages of this project. This work is partially supported by CONICET grant PIP-11220110100005 and PICT-2016-1358.

\end{document}